\documentstyle[12pt,epsfig]{article}
\newcommand{\be}{\begin{equation}}
\newcommand{\ee}{\end{equation}}
\newcommand{\ba}{\begin{eqnarray}}
\newcommand{\ea}{\end{eqnarray}}
\newcommand{\ci}[1]{\cite{#1}}
\newcommand{\lab}[1]{\label{#1}}
\begin{document}

\title{  Determination of the structure \\
   of the high energy hadron elastic scattering amplitude at small angles}

\author{ B. Nicolescu 
  \\
 Theory Group,  Laboratoire de Physique Nucl\'eaire  et des Hautes   \\
Energies (LPNHE)\thanks{Unit\'e  de Recherche des Universit\'es 
  Paris 6 et Paris 7, Associ\'ee ou CNRS} ,
 CNRS and Universit\'e Pierre et Marie Curie, Paris, \\
     e-mail: nicolesc@lpnhep.in2p3.fr \\
\phantom{} \\
  O.V. \ Selyugin   \\
 BLTP, JINR and Universit\'e  de Li\`ege   \\
    email: selugin@thsun1.jinr.ru }




\maketitle

\abstract
 { A new method for the determination of the real part of the elastic scattering
  amplitude  is examined for high energy proton-proton and proton-nuclei 
   elastic scattering at small momentum transfer. This method allows us 
   to decrease the number of  model assumptions,  to obtain the real part  
   in a narrow region  of momentum transfer    
    and to test  different models
   for hadron-nuclei scattering. 
}


\vspace{0.5cm}

\section{Introduction}

     The investigation of the structure of the hadron scattering amplitude
 is an important task both for  theory and  experiment.
 PQCD cannot   calculate neither
 the value of the scattering amplitude,
 nor its phase or its  energy dependence
 in the soft diffraction range.
 A  deeper understanding of the way  that 
 such  fundamental relation as integral dispersion and
 local dispersion relations  work requires the knowledge of 
 the structure of the
 scattering amplitude with high accuracy \cite{mart}.
 It was shown in \ci{kh}  that the knowledge of the behavior
 of $\rho$ - the    ratio of the real to the imaginary part of
 the spin-non-flip amplitude - 
  can be used for checking   local quantum field theory (QFT)
    already in the LHC energy region.

      A large number of experimental and theoretical studies of 
 high-energy elastic
 proton-proton  and proton-antiproton scattering at  small angles
 gives a rich information about this  process,
 and  allows to narrow the circle of examined models
 and  to  point to a number of  difficult  problems
 which are not yet solved  entirely.
 This concerns especially  the  energy  dependence  of  a number
 of characteristics of these reactions and the contribution of the odderon.

    Many of these questions are connected with the dependence 
 with $s$ and $t$
 of the spin-non-flip phase of hadron-hadron scattering.
 Most of the models define  the  real  part  of  the  scattering
 amplitude phenomenologically. Some  models  use  the  local
 dispersion relations       and   the   hypothesis   of  geometrical
 scaling. As is well known, using some  simplifying assumption, the
 information about   the   phase   of
 the   scattering   amplitude   can   be
 obtained from the experimental data at small momentum transfers   where
 the interference of the electromagnetic and hadronic amplitudes takes  place.
 On the whole, the obtained information confirms the local dispersion
 relations.
   It was shown in \ci{k-l} that a self-consistent
 description of the experimental data in   the energy range
 of the ISR and the SPS  can be obtained in the case of a rapidly  changing
 phase when the real part of the scattering amplitude grows quickly at 
  small t   and   becomes dominant. 
   
   Now the physics of high-energy nuclei scattering is developing quickly
   and 
  the knowledge of the structure of the elastic proton-nuclei and
  nucleus-nucleus scattering is needed to discriminate between
  different models
  describing high-energy nuclei interactions. This is  especially  important 
  in view of the development of the QCD approach 
 to the high-energy nuclei interaction  \ci{armesto}.

     The standard procedure to extract the magnitude of the real part
  includes a fit to the experimental data taking the magnitude
   of the total hadronic cross section, the slope, $\rho$, and,
  sometimes the normalization  coefficient corresponding to luminosity
  as free parameters:
\ba
   \sum_{i}^{k} \frac{(n d\sigma^{exp}/ dt(t=t_i)
                -d\sigma/ dt(t=t_i) )^2}{\Delta^{2}_{exp,i}} \lab{fit}
\ea
where $ d\sigma^{exp}/ dt(t=t_i)$ is the differential cross sections
 at point $t_i$,  with the statistical error $\Delta_{exp,i}$
extracted from
 the measured $dN/dt$ using, for example, the value of the luminosity,
 $n$ is a fitted normalization parameter.
This procedure requires a sufficiently wide interval of $t$ and a large number
 of experimental points.

   The theoretical representation of the differential cross-sections is
\ba
{d\sigma\over dt}&= 2 \pi
[\vert \Phi_1\vert^2+\vert \Phi_2\vert^2
    + \vert \Phi_3\vert^2+\vert \Phi_4\vert^2+4\vert
 \Phi_5\vert^2]\ ,
\ea
 where  $\Phi_1$ and $Phi_3$ are the spin non-flip amplitudes.
The total helicity amplitudes can be written 
 as a sum of nuclear $\Phi_i^h(s,t)$
and electromagnetic  $\Phi_i^e(s,t)$ amplitudes :
\ba
\Phi_i(s,t)=\Phi_i^h(s,t)+e^{i\alpha\varphi}\Phi_i^e(t)\ ,
\ea
where $\Phi_i^e(t)$ are the leading terms at high energies for the one-photon
amplitudes as defined, for example,
 in \ci{leader} and $\alpha$ is the fine-structure constant. 
  The common phase $\varphi$      is 
\ba
\varphi= \pm[\gamma+\log\big(B(s,t)\vert t\vert/2\big)+\nu_1+\nu_2],
\ea
where the upper(low) sign related to the $p\bar{p}$($pp$) scattering, and
$B(s,t)$ is the slope of the nuclear amplitude, 
 and $\nu_1$ and $\nu_2$ are
small correction terms  defining the behavior of the Coulomb-hadron phase
at small momentum transfers (see, \cite{Cahn} or more recently \ci{selprd}).
At very small $t$ and fixed $s$, these electromagnetic amplitudes 
 are such that
$
\Phi_1^e(t) = \Phi_3^e(t)\sim\alpha/t \ ,
\Phi_2^e(t) = -\Phi_4^e(t)\sim \alpha\cdot \hbox{const.}\ ,
\Phi_5^e(t)  \sim  -\alpha/\sqrt{\vert t\vert}\ .
$
  We assume, as usual, that at high energies and small angles
  the double-flip amplitudes are small with respect to the spin-nonflip one
  and that spin-nonflip amplitudes are approximately equal. Consequently,
  the observables  are determined by two spin non-flip  amplitudes:
  $ F (s,t) = \Phi_{1}(s,t) + \Phi_{3}(s,t)
   =F_{N}+F_{C} \ \exp( i \alpha \varphi)$.

   In the case of high-energy hadron scattering, we can neglect
   the contribution of the spin-flip amplitude at small momentum transfer 
   in  the differential cross section and write
 in the $O(\alpha)$ approximation
:
\ba
     d\sigma/dt = \pi |e^{i \alpha \varphi} F_C + F_N|^2 =
                    \pi \ [ (  F_C + Re F_N)^2
               +(\alpha \varphi F_C + Im F_N)^2] \, .  \lab{ds0}
\ea

    In the standard fitting procedure, this equation takes the form:
\ba
d\sigma/dt &= \pi [ (F_{C} (t))^2
          + (\rho(s,t)^2 + 1) (Im F_{N}(s,t))^{2})      \nonumber \\
 &+ 2 (\rho(s,t)+ \alpha \varphi(t)) F_{C}(t) Im F_{N}(s,t)], \label{ds2}
\ea
where $F_{C}(t) = \mp 2 \alpha G^{2}/|t|$ is the Coulomb amplitude
(the upper sign is for $pp$, the lower sign is for $p\bar{p}$)
 and $G^{2}(t)$ is  the  proton
electromagnetic form factor squared;
$Re\ F_{N}(s,t)$ and $ Im\ F_{N}(s,t)$ are the real and
imaginary parts of the nuclear amplitude;
$\rho(s,t) = Re \ F_{N}(s,t) / Im \ F_{N}(s,t)$.
The formula (\ref{ds2}) is used for the fit  of  experimental  data
determining the Coulomb and hadron amplitudes and the Coulomb-hadron
phase to obtain the value of $\rho(s,t)$.

\section{The real part of the spin-non-flip amplitude of the $pp$ scattering} 

   Numerous  discussions of the function $\rho (s,t)$
 measured by
 the UA4 \ci{ua4} and UA4/2 \ci{ua42} Collaborations at $\sqrt{s}=541$ GeV 
 have revealed
 the ambiguity in the definition of this semi-theoretical parameter \ci{selpl},
 and, as a result, it has
 been shown that one has some trouble in extracting, from experiment,
  the total cross sections
 and the value of the forward ($t=0$) real part of the scattering 
 amplitudes \ci{gns}.
 In fact, the problem is that we have at
  our disposal only one observable $d\sigma/dt$
 for two unknowns,  the real  and imaginary parts of $  F_{N}(s,t)$.
 So, we need either some additional experimental information which
 would allow us to determine independently the real and imaginary parts of
 the spin non-flip hadron elastic scattering amplitude or some new ways
  to determine the magnitude of the phase of the scattering amplitude
  with minimum theoretical assumptions.
          One of the most important points in the definition
  of  the real part of the scattering amplitude
  is the knowledge of the normalization coefficient and of
  the magnitude of $\sigma_{tot}(s)$.

     To obtain the magnitude of   $Re F_{N}(s,t)$,
  we fit the differential cross sections taking into account 
   the value of $\sigma_{tot}$ either
  from another experiment, to decrease the
 errors,  as made
 by the UA4/2 Collaboration, or   as a free
 parameter, as done in \ci{selpl}.
 If one does not take the normalization coefficient as  a free parameter in
 the fitting procedure, its definition requires the knowledge of
 the behavior of the imaginary and real parts of the scattering amplitude
 in the range of small momentum transfer, of the magnitude of
 $\sigma_{tot}(s)$ and of $\rho(s,t)$.

   Let us note three points. First, we should take into account
 the errors on $\sigma_{tot}(s)$. Second, this method implies that
 the slope of
 imaginary part  of the scattering amplitude is equal to the
 slope of its real part in  the examined range of momentum transfer,
 and, for the best fit, one should take the interval of momentum transfer 
  sufficiently large.
 Third, the magnitude of $\rho(s,t)$ thus obtained  corresponds to
 the whole interval of momentum transfer .

     In this article, we briefly describe   new procedures  simplifying
 the determination of elastic scattering amplitude parameters.

   From equation   (\ref{ds0}), 
    one can obtain an equation for  $Re F_{N}(s,t)$
   for every experimental point  $i$: 
\ba
  && Re F_{N}(s,t_i)= - F_{C}(t)   \nonumber    \\
  & & \pm [(1+\delta) / \pi d\sigma^{exp}/ dt(t=t_i)
   - (\alpha \varphi F_{C}(t_i)+Im F_{N}(s,t_i))^2]^{1/2},
                                               \label{rsq}
\ea
where
\ba
  \delta = \epsilon/N, \lab{delta}. 
\ea
  $N$ in eq.  (\ref{delta})
   being the normalization constant in a well-defined experiment 
   by using a particular method of obtaining 
  the normalization (for example,  by luminocity) 
  and $\epsilon$ - the inevitable error on $N$
  due to the use of a particular method.
  For example, in method of luminosity it reflects the error in the mesuare of
  luminosity.
   Therefore $n$ from eq. (\ref{fit}) is given by 
\ba
n=1+\delta,
\ea

 We define the imaginary part of the scattering amplitude {\it via} the
 usual exponential approximation in the small $t$-region
\be
  Im F_{N}(s,t) = \sigma_{tot}/(0.389 \cdot 4 \pi) \exp(B t/2), \lab{im}
\ee
 where $0.389$ is the usual converting dimensional factor for expresing 
 $\sigma_{tot}$ in mb.
 
  It is evident from  (\ref{rsq}) that the determination of
 the real part depends on
 $\delta, \sigma_{tot}, $  and $B$,
     the magnitude of $\sigma_{tot}$ depending itself on $\delta$.  
     Equation (\ref{rsq}) shows  the possibility to calculate
  the real part at every separate point $t_i$ if the imaginary part of
  the  scattering amplitude and $\delta$ are fixed,
  and to check the exponential form
  of the obtained  real part of the scattering amplitude 
  (see \cite{selyf}).

   Let us define the sum of the real parts of the
  hadron and Coulomb amplitudes 
  as $\sqrt{\Delta_{R}}$, so we can write:
\ba
  \Delta^{th.}_{R}(s,t_i) =  [ Re F_{N}(s,t_i)+ F_{C}(t)]^2. \label{Del}   
\ea
 Using  the experimental data on the differential 
 cross sections we obtain:
\ba
  \Delta_{R}(s,t_i)&=& \Delta^{exp.}_{R}(s,t_i) =  \nonumber \\
   ( 1 + \delta ) / \pi \  d\sigma^{exp.}/ dt(t&=&t_{i}) 
           - ( \alpha \varphi F_{C}(t_i)+Im F_{N}(s,t_i))^2 
  \label{Del2}
\ea

        This formula has a  significant property for
  the proton-proton scattering at  very high energy,
 but, of course,  non-asymptotic,   is sufficiently large and  
 opposite in sign  relative to the Coulomb amplitude
 and for proton-antiproton  scattering at low energy
 where the real part of the hadron amplitude is negative.
 Let us put the representation of the differential cross sections, 
 using  eg.(\ref{ds2}),  in eq. (\ref{rsq}), taking into account
  that we do not know  exactly the normalization of the differential 
 cross sections.
  So,  we get
 \ba
  && \Delta_{R}(s,t_i) =  (1+\delta)( Re F_{N}(s,t_i) +  F_{C}(t))^2 
        \nonumber  \\
  & &  
   + \delta (\alpha \varphi F_{C}(t_i)+Im F_{N}(s,t_i))^2.
                                               \label{Del1}
\ea
  As expected,  
  in the standard picture of high-energy hadron scattering at small momentum 
  transfer,   the real part of $F_{N}(s,t)$ 
  is positive and non-negligible in the region 
   $50 \ $~GeV $ \leq \sqrt{s} \leq 20 \ $TeV.
    The experiments at  $\sqrt{s} \simeq 50 \ $ GeV for
  proton-proton and at  $541 \ $GeV for proton-antiproton scattering
  support this picture.
  Hence, using the experimental data of the differential cross sections 
 on high energy $pp$-scattering  at some energy $s_j$  and the  
 the  imaginary part of the hadron amplitude, we can calculate the value 
  $\Delta_R(s_j, t_i)$.

      Let us examine this expression for the $pp$ scattering amplitude
  at energies
      above $\sqrt{s} = 500 \ $GeV.
  In order to do this, let us make a gedanken experiment and calculate
  $d\sigma/dt_i$ at some high energy. 
  For that let us take the hadron amplitude in the exponentially form
  with fixed parameters ($\rho= 0.15$ and $\sigma_{tot}=63 \ $mb) 
  and calculate the differential cross sections in some number of
  points of $t$. These values  will be 
  considered as ``experimental'' points at  $d\sigma_i/dt$ 
   of the differential cross section  at  $t_i$.
  In this case, we exactly
know the all parameters of hadron amplitude from beginning
 and can check the our final result by  compare it with the input parameters.

  For  $pp$ scattering at high energies,
  equation (\ref{Del1})
   induce  a remarkable property:
  the real part of the Coulomb $pp$ scattering amplitude 
  is negative and
  exceeds the size of $F_N (s,t)$ at 
  $t \rightarrow 0$ , but it has a large slope.
   As the real part of the hadron amplitude is positive at high energies,
    it is obvious
 that  $\Delta_R$ has a minimum at a position in 
  $t$  independent of $n$ and of  $\sigma_{tot}$, as shown
 in Fig. 1.

  The position of the minimum  gives us the value 
  $t=t_{min}$ where $Re F_{N} = - F_{C}$. As we know the
 Coulomb amplitude, we can estimate the real part of the
 $pp$ scattering amplitude at this point. 
  Note that all other methods give us
 the real part only in a rather  wide interval of momentum transfer.
The true  normalization coefficient and 
 the true value of $\sigma_{tot}$ will correspond to a zero in 
 $\Delta_R(s,t)$ at the point $t_{min}$.
 But if the normalization coefficient
  is not the true one, 
  the minimum will be or above or below zero, but practically 
  at the same point $t_{min}$. So, the size of $\Delta_R$ also tests the 
  validity of the determination of the normalization coefficient 
  and of  $\sigma_{tot}$.

  This method works only in the case of  positive real part
 of the hadron amplitude and it is especially
  efficient in the case of large $\rho$. So, this method is
 interesting for the  experiments which will be done
  at RHIC and for the future TOTEM experiment   at LHC.

    In spite of the fact that at ISR energies 
 we have small $\rho(s, t \approx 0)$ and
 few experimental points, let us try to examine one experiment,
 for example, at $\sqrt{s}=52.8 \ $GeV. This analysis is shown in Fig.2.
 One can see that in this case the minimum is sufficiently large, and
 $-t_{min}= (3.3 \pm 0.1)10^{-2} \ $GeV$^2$.
 The corresponding real part
 is equal to  $0.38 \pm 0.014 \ $mb$^{1/2}$/GeV. 
 Our analysis gives
 $\rho(t_{min}) = 0.054 \pm 0.003$, as compared with 
 $\rho(t=0) = 0.077 \pm 0.009$  as given in \ci{528}.

  Now let us see what we can obtain in the future experiment
  on the proton-proton elastic scattering at maximum energy of RHIC.
  For that
  we can simulate the experimental data 
on the differential cross sections by  calculating with
  some model for the imaginary and  real parts of the hadron amplitude
  with definitely parameters $\sigma_{tot}, B, \rho$ as we made above.
  After that we have to taking account the possible statistic errors
   expected in  future experiments.
   Namely we calculate the deviation
  from the theoretical values of the differential cross sections
  (in  units of error bias)  at examined point of $t_i$
   by using a Gaussian random
  procedure  to calculate  the probability of the deviation.
  After that we change our theoretical differential cross section 
  on this deviation. In result we obtain the ``simulated experimental'' data
  with some ``experimental'' errors. These ``experimental'' data
   will  have the Gaussian distribution   from the theoretical curve.
  As a  result, we can simulate the future experimental data
  for the differential cross sections,
  for example,  with the possible $\rho$ values  
  $0.135$ or  $0.175$.
  Then we can get the values of $\Delta_R$ from
   these  gedanken `` experimental'' data 
  with correspounding statistical errors. 
   These values are shown 
    in Figs. 3a and 3b and we can note  
   the difference between the two  respective
    models for the ``data'' corresponding to two different values of $\rho$. 
   The pure theoretical representation of $\Delta_R$ 
   with   $\rho=0$ and the same values of $\rho$ as above 
    are also shown. 
   
  There are other two interesting features:
  the magnitude and the position of the second maximum.
   It is easy to  connect the size of the maximum with the magnitude of the
  real part of the scattering amplitude.
  Let us consider the  $t$-region very near the minimum or the
  maximum in  $\Delta_R$. In this region, we can approximate 
  the $t$-dependence of the electromagnetic form factor by 
  an exponential of slope $D$ and $Re F_N (s,t)$ by an exponential
  with slope $B_r$. We  then equate the derivative of $\Delta_R$
   to zero:
\ba
\frac{d}{ dt}[  \Delta_R] = \frac{d}{ dt}[ h_1^2\frac{1}{ t^2}
 e^{ D t}
   + 2 h_1 h_2 \frac{1}{ t}e^{ Dt/2} e^{ B_r t/2} +  
   h_2^2 e^{B_r t}] =0, 
\ea
 where $h_1$ and $h_2$ are some electromagnetic and hadronic constants.
 
 Therefore, at  $-t=t_{max}$, where  $t_{max}$  
  is corresponding to the second  maximum of 
  $\Delta_R$, we get
\ba
 &&  2 h_1^2 e^{-D t_{max}}  + h_1^2 D t_{max} e^{-D t_{max}}
   - 2 h_1 h_2 \frac{D+B_r }{ 2}  t^2_{max} e^{-Dt_{max}/2}  
      e^{-B_r t_{max}/2}  \nonumber \\
 &&   -2 h_1 h_2 t_{max} e^{-D t_{max}/2}  e^{-B_r t_{max}/2} 
   +  h_2^2  t_{max}^3 B_r    e^{-B_r t_{max}}  =0.  
\ea 
  
 No term in this equation can be neglected, because in the region
  of interest 
 all these terms are of  the same order of magnitude.
As a  result, we obtain  a simple quadratic equation  at  $-t=t_{max}$:
\ba
\frac{ Re F_N^2}{ Re F_C^2} \  \frac{ B_r t_{max}}{ 2} 
 + \frac{ Re F_N}{ Re F_C}  (1+\frac{D+B_r }{ 2}  t_{max}) 
    + \frac{D}{ 2}  t_{max} +1 = 0. 
\ea
 It  leads to the  simple relation
\ba
   B_r /2 =  (1+\frac{D}{ 2} t_{max} ) \ 
                 \frac{-1}{t_{max}} \frac{  F_C}{  Re F_N}.
\ea

  Remembering the definition of $\Delta_R$, we obtain
\ba
 B_r/2 =       (1+\frac{D}{ 2} t_{max} )  \ 
   \frac{ 1}{ t_{max}} \frac{1}{ ( 1 - \Delta_{R}^{ 1/2} /Re F_C )}.
\ea

  So, we can determine the slope of the real part of the hadron 
 elastic scattering amplitude without any fitting procedure in a large
  interval of momentum transfer.


   Note that the point  $t_{min}$ is also
  important for
   the determination of the real part of the spin-flip amplitude  
  \cite{PS1}.
  At that point, some terms 
 in the definition of the analyzing power
  will be canceled. 
  Together with them from the basic equation the imaginary part of 
  the spin-flip amplitude disappear also.
 Such
  reducing
representation can be used
  for the determination of the real part of the hadron spin-flip
  amplitude at high energy and small angles.

\section{The real part of the spin-non-flip amplitude of the $pA$ scattering} 

   It is interesting to apply this new method to  proton-nucleus
   scattering at high energies. 
    The size of the hadron amplitude grows only slightly
   less then  $A$,
    the atomic number: for example, $\sigma_{tot}(pp) = 38 \ $mb
   and   $\sigma_{tot}(p ^{12}C)= 335 \ $mb   in the region of $100 \ $GeV. 
   The most important difference 
   in proton-nucleus scattering as compared with  
  $pp$ scattering is that the slope 
  of the differential cross section 
 is very high:  $\simeq 70 \ $GeV$^{-2}$
   for  this nuclear reaction at $100 \ $GeV.
  The electromagnetic amplitude grows like $Z$. 
  Its slope also grows. It is interesting that the simple calculations,
  which take the hadron amplitude  at small momentum transfer 
  in the usual exponential form with a
  large slope, lead to  practically the same results as for 
  proton-proton scattering.

   Let us take the Coulomb amplitude for $p ^{12}C$ scattering in the form
\ba
 F_{C}&=& {2 \alpha_{em} \ Z\over t} \ F^{^{12}C}_{em} F^{p}_{em1}
                        F^{^{12}C}_{em} F^{p}_{em2},
\ea
where 
$ F^{p}_{em1}$ and $ F^{p}_{em2}$ are
the electromagnetic form factors of the proton, and
$F^{^{12}C}_{em}$ that of $^{12}C$.
We use 
\ba 
F^{p}_{em1}&=& {4 m_p^2-t (\kappa_p+1)\over (4 m_p^2-t)(1-t/0.71)^2} , \\ 
F^{p}_{em2}&=& {4 m_p^2\kappa_p\over (4 m_p^2-t)(1-t/0.71)^2} , 
\ea
 where $m_p$ is the mass of the proton and $\kappa_p$ the proton  anomalous 
magnetic moment.
We obtain $F^{^{12}C}_{em}$
from the electromagnetic density of the nucleus \cite{jansen}
\ba
D(r) = D_0 \ \left[1+ \tilde{\alpha}\left({r\over a}\right)^2
\right]e^{-\left({r\over a}\right)^2},
\ea
$\tilde{\alpha}=1.07$ and $a=1.7$ fm giving the best description
 of the data  in the small $t$-region
and producing a zero of $F^{^{12}C}_{em}$ at $|t| = 0.130 $~GeV$^2$. 
We also calculated $F^{^{12}C}_{em}$ by integrating  the nuclear 
form factor as given by a sum of Gaussians \cite{data35} and we obtain 
practically the same result, the zero being now at  $|t| = 0.133$ 
GeV$^2$. 

   We take the hadron amplitude in the standard exponential form
  with the parameters obtained in \cite{SELEX}, 
   $\sigma_{tot}=335 \ $mb and $B=62 \ $GeV$^{-2}$.
  The calculations shown in  Fig. 4 for two variants (with $\rho= 0.1$
  and  $\rho =0.075$) demonstrate that
  the  minimum is situated 
   approximately in  the same $t$-region as for the minimum in 
  proton-proton scattering. There is also a  significant difference in the
  size of the maximum for these two values of $\rho$,
 which is connected with the large 
  slope of  proton-nucleus  scattering.

   Such calculations were also carried out for $p ^{28}$Si reaction.
  In this case, we determine the 
   electromagnetic form-factor as sum of gaussians
  in the $r$-representation  \cite{data35}.
   The parameters of the hadron scattering amplitude were chosen
  near the parameters for $p ^{27}Al$ scattering, given by \cite{shiz}:
  $\sigma_{tot} = 800 \ $mb and $B= 120 \ $GeV$^{-2}$. 
  The results are shown in Fig.5. It is clear that we obtain a very similar 
  situation, weakly dependent on the specific nucleus.

   All previous results were obtained under
   the assumption that hadron scattering
   amplitude has the same exponential behavior 
   as the $pp$ scattering amplitude at high energies. Very frequently
   the Glauber model  is used for the description of hadron-nuclei reactions
    and
   this model gives a different behavior for the  hadron scattering amplitude
   at small momentum transfer.
 The slope of the hadron  amplitude  of the elastic proton-nuclear 
 scattering increases
  with  $|t|$ (see Fig.6),  like in ``black body'' limits.  
 But   in $pp$ scattering, the slope is either 
    constant or slightly decreasing with  $|t|$ at small momentum transfer
  region .
    At low energy a good description was obtained 
    for different nuclear reactions in the framework 
    of Glauber model \cite{GM}.
   Note that  proton-proton scattering at low energy is also predicted 
   by the Glauber model to have   the same behavior
     as  nuclear reactions  at low transfer momentum.
    In Fig.6, the slopes of the real and imaginary part of the hadron
   amplitude of $p ^{12}C$ elastic scattering 
 calculated in the Glauber model 
 are shown.
  
     The calculations were obtained by 
  the formulas used in Refs. \cite{gl0} and  \cite{k-t}. 
    It is clear that the real part 
   decreases very fast  and changes sign at $-t= 0.06  \ $GeV$^2$.
    $\Delta_R$ with $\rho = 0.1$ has a wide
    minimum in the region  $0.025 \leq |t| \leq 0.045 \ $GeV$^2$
     (see Figs. 8a and 8b).
    But if we perform such a calculation for $\rho = -0.1$,
   the ordinary minimum
   is obtained but is situated at a large value of $t$. 
   This can be understood
   from Fig.7, were we  show the real parts of the hadron amplitude
    (for $\rho= \pm 0.1$) and of the Coulomb amplitude. In the region 
    around $|t| \simeq 0.05 \ $GeV$^2$  the slopes of the two amplitudes
    coincide. A significant cancellation occurs for $\rho = 0.1$.
    All these results come from the behavior of the slope of the hadron
     amplitude in the Glauber model.
    As a result, we obtain a very different behavior of $\Delta_R$
    in the Glauber model as compared  with the exponential behavior.
    So, the investigation of $\Delta_R$ can distinguish
    different approaches.   

\section{Conclusion}

 The precise experimental measurements of $dN/dt$ and 
  the spin correlation parameter $A_N$ at
 RHIC, as well as, if possible, at the Tevatron, will therefore give us
 unavailable information on  hadron elastic scattering at small t. New
 phenomena at high energies \cite{osc}
  could be therefore detected without going 
 through the usual arbitrary assumptions (such as the exponential form) 
 concerning the behavior of the hadron elastic scattering amplitude. 
  It is  interesting to apply this method to  proton-nuclei
  scattering at high energy, especially at RHIC energies. It gives
  a unique possibility to investigate  the real part 
  of the hadron amplitude in  nuclear reactions.

\newpage

     {\it Acknowledgment.} 
 The authors express their thanks to J.-R.~Cudell, J. Cugnon, W. Guryn and 
 E. Martynov 
for fruitful discussions. One of us (O. S.) thanks Prof. J.-E. Augustin 
for the hospitality at the LPNHE Paris, where part of this work was done.

\newpage
     Captions

FIG.1.
  $\Delta_R$
 for   $pp$ scattering
 at $\sqrt{s}=540 \ $GeV  and with $\sigma_{tot} = 63 \ $mb,
 for  different $n$ [triangles - the calculations by (\ref{Del2});
  curves and circles - the calculations by (\ref{Del1})].

FIG.2.
 $\Delta_R$ for  $pp$ scattering
 using the experimental data for $d\sigma/dt$ at $\sqrt{s}=52.8 \ GeV$
  \cite{528}. The lines are  polynomial fits to the points calculated
  with experimental data and with different $n$

FIG.3 (a,b). 
 $\Delta_R$ for  $pp$ scattering
 at $\sqrt{s}=540 \ $GeV  with a) $\rho_1 =0.135$ and b) $\rho_2 =0.175$.
 The solid, dashed, and dotted lines are the
 theoretical curves calculated by eq. (\ref{Del})
  for $\rho_2 = 0.175$, $\rho_1 = 0.135$ and $\rho_0 = 0$
  respectively. The points show the  $\Delta_R$ calculated from
  the ``simulated experimental'' data $d\sigma/dt$ for both cases.

FIG.4.
   $\Delta_R$ for  $p ^{12}C$ scattering
   with  $\rho =0.1$ and  $\rho =0.075$ 
 (  solid and  dashed lines respectively) 
  for the exponential behavior of the hadron amplitude.

FIG.5.
  $\Delta_R$ for  $p ^{28}Si$ scattering
   with  $\rho =0.1$ and  $\rho =0.075$ 
 ( solid and  dashed lines respectively) 
  for an  exponential behavior of the hadron amplitude.

FIG.6.
  The  slope $B_{gl}$ of the real (hard line) and imaginary (dashed line) parts
   of the hadronic amplitude, calculated from  the Glauber model for
  $p ^{12}C$ scattering. 

FIG.7.
   The real part of the electromagnetic  amplitude (hard line) and
  the real part of the  hadron amplitude (dashed line)  
     calculated from the Glauber model for
  $p ^{12}C$scattering with $\rho= 0.1$ (long-dashed line) and with
  $\rho=-0.1$.  
  . 

FIG.8 (a,b).
  $\Delta_R$ for  $p ^{12}C$ scattering
   with  $\rho =0.1$ and  $\rho = - 0.1$ 
 ( solid and  dashed lines respectively ) calculated   
  in the framework of the Glauber model, 
  with  linear (a) and logarithmic (b) scales.

\newpage

\begin{figure}[!ht]
\epsfysize=90mm
\centerline{\epsfbox{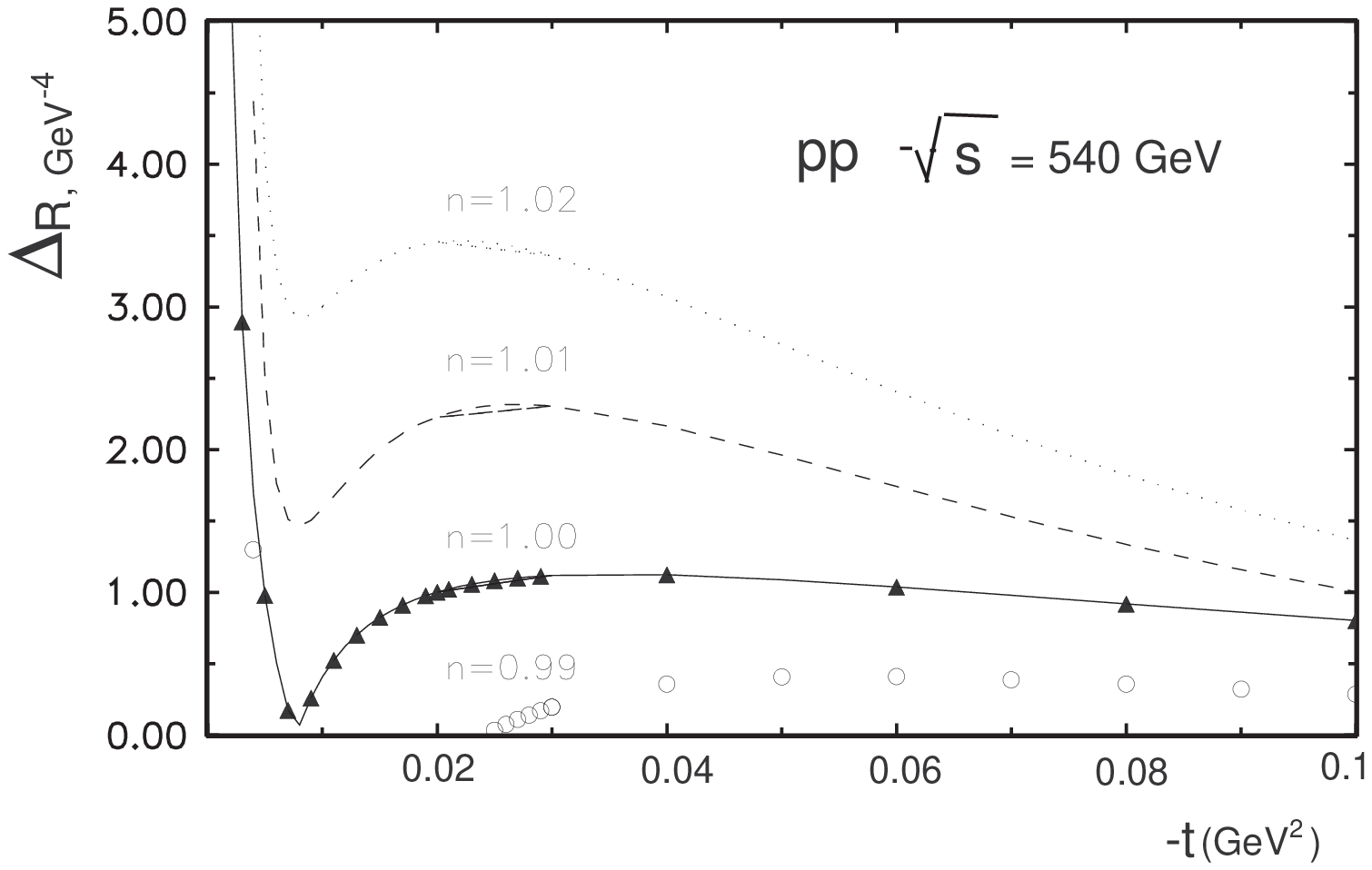}}
\vspace{-0.5cm}
\centerline{Fig.1 }
\epsfysize=80mm
\centerline{\epsfbox{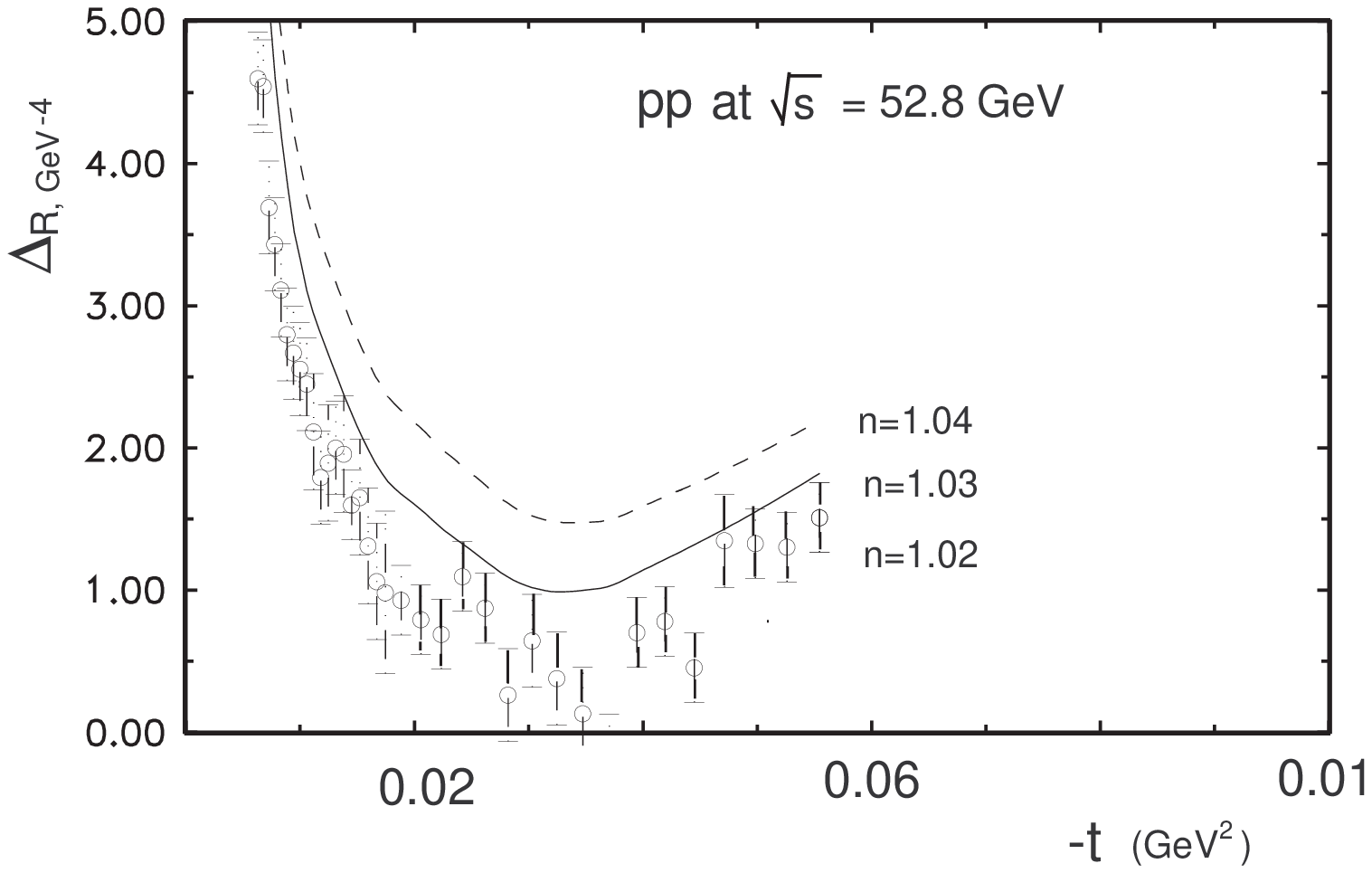}}
\vspace{-0.5cm}
\centerline{Fig. 2}
\end{figure}


\newpage

\begin{figure}[!ht]
\epsfysize=80mm
\centerline{\epsfbox{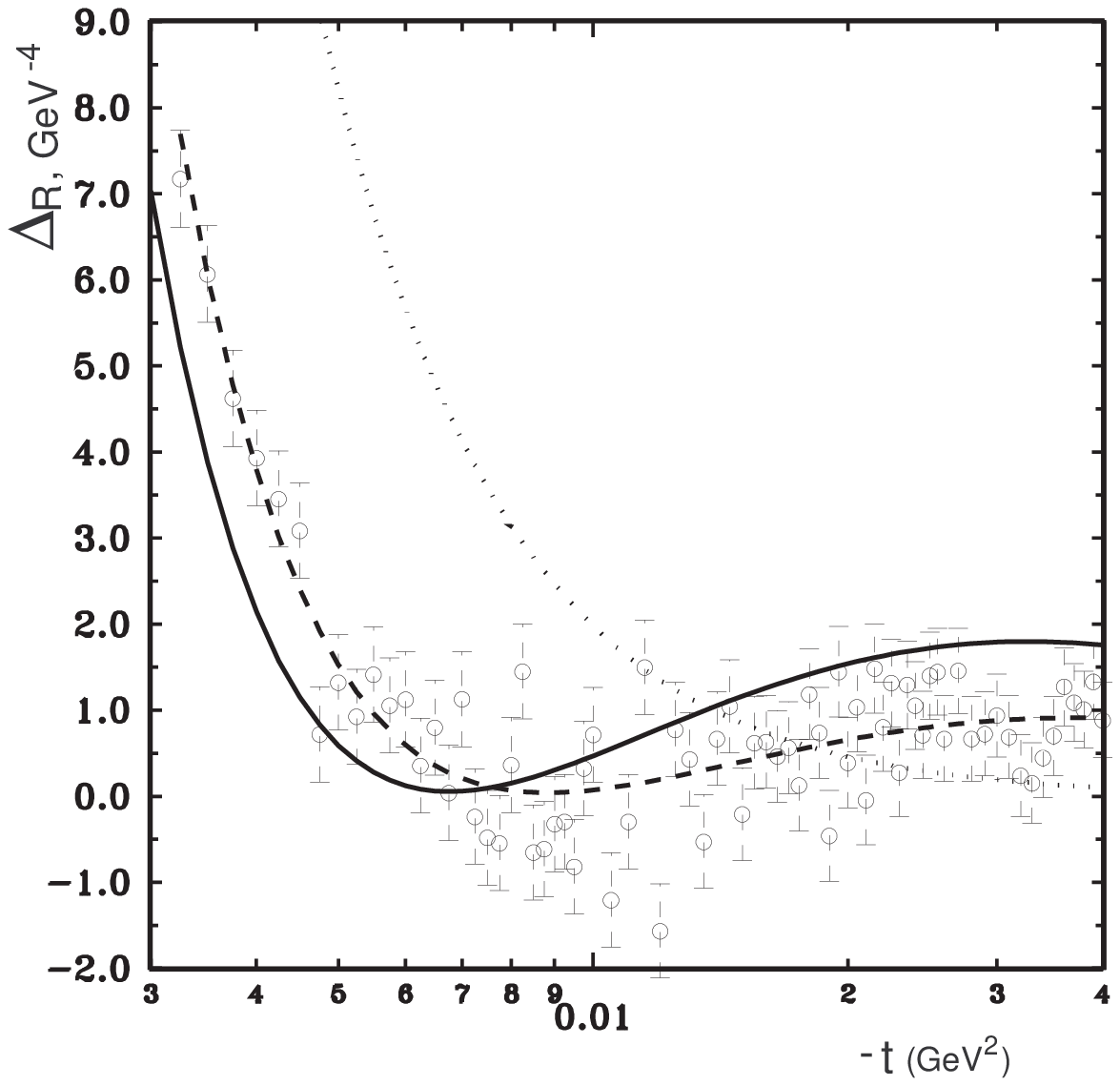}}
\vspace{-0.5cm}
\centerline{Fig.3 a}
\epsfysize=80mm
\centerline{\epsfbox{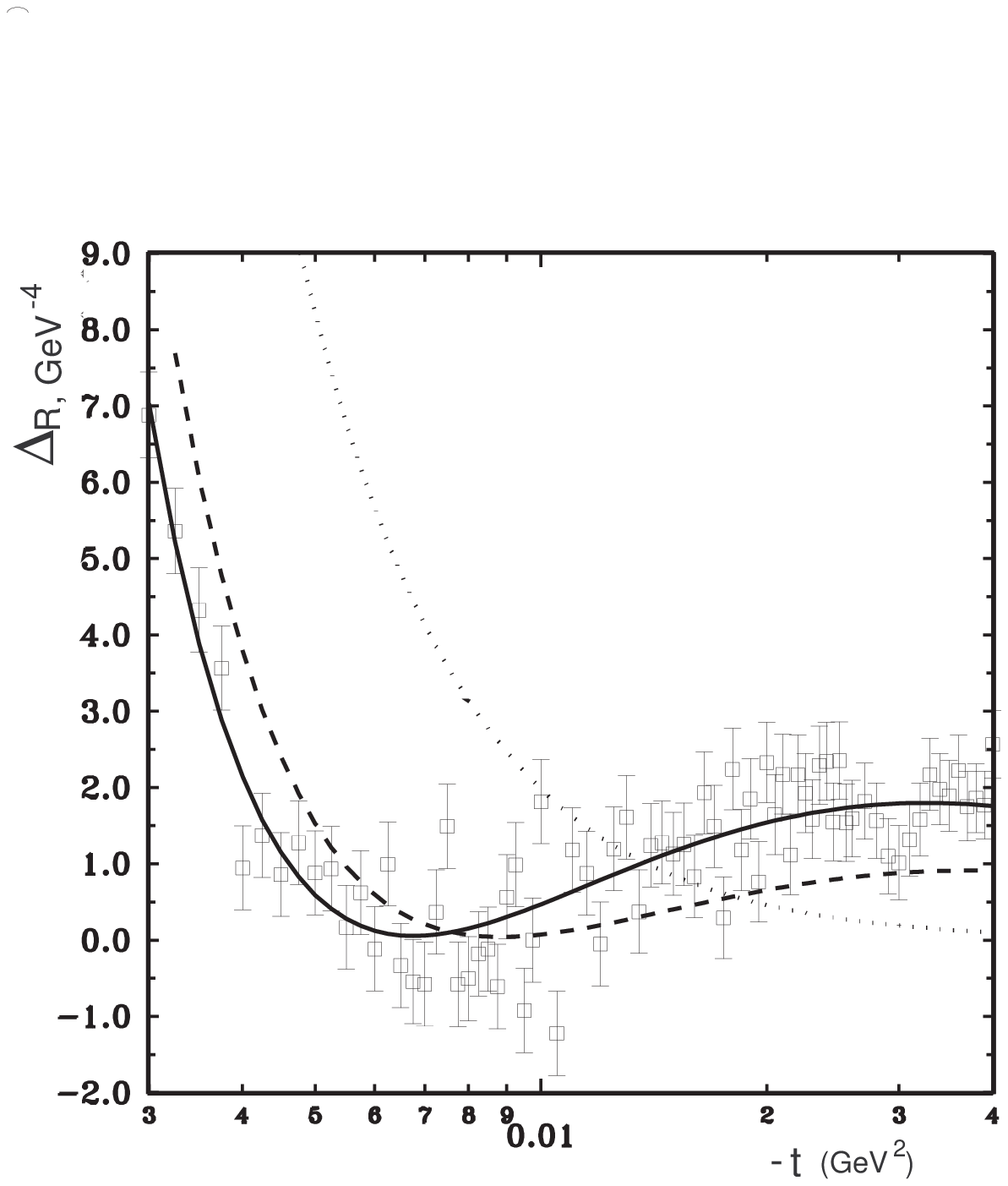}}
\vspace{-0.5cm}
\centerline{Fig.3 b}
\end{figure}

\newpage

\begin{figure}[!ht]
\epsfysize=80mm
\centerline{\epsfbox{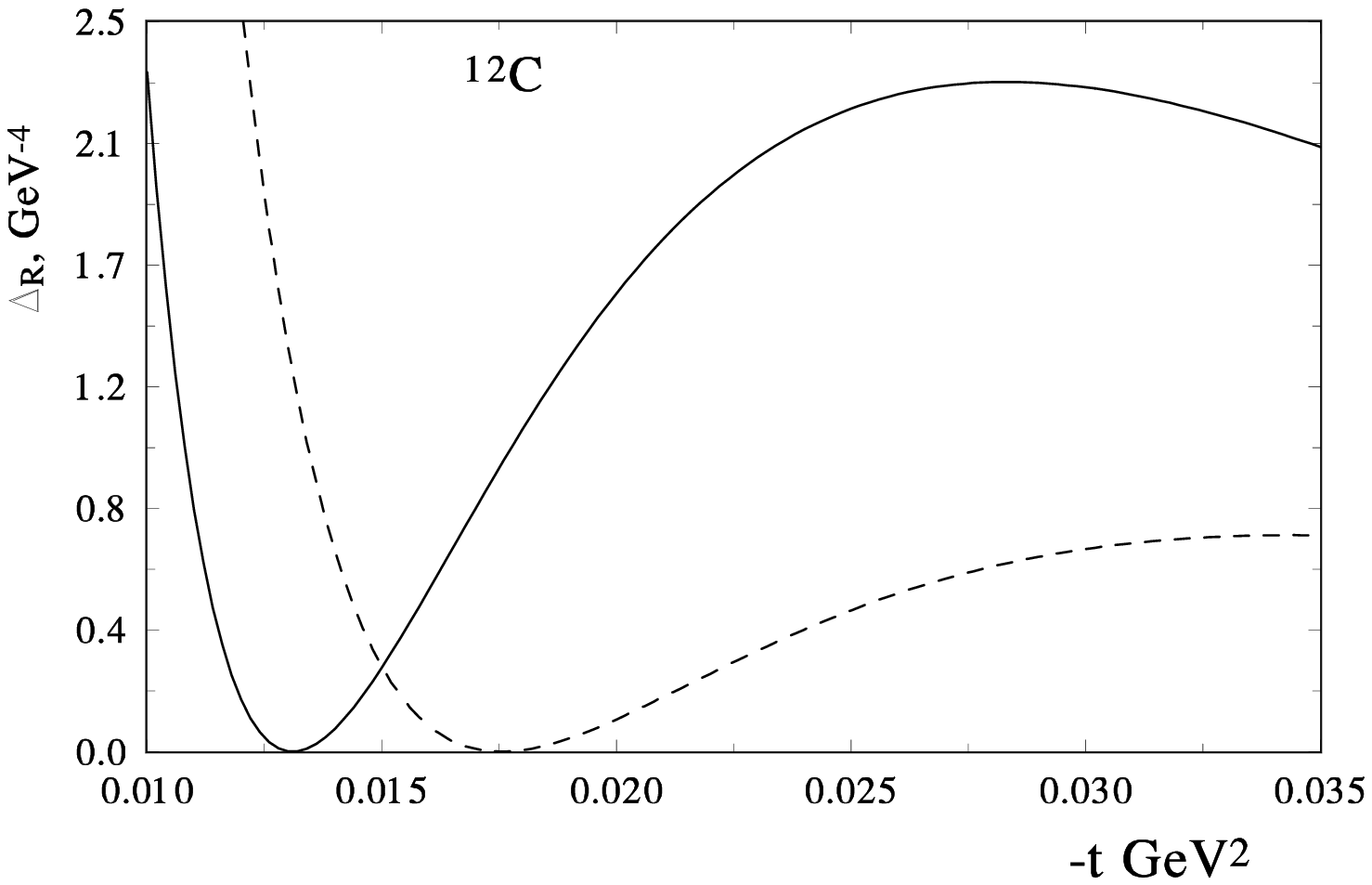}}
\vspace{-0.5cm}
\centerline{Fig.4 }
\epsfysize=80mm
\centerline{\epsfbox{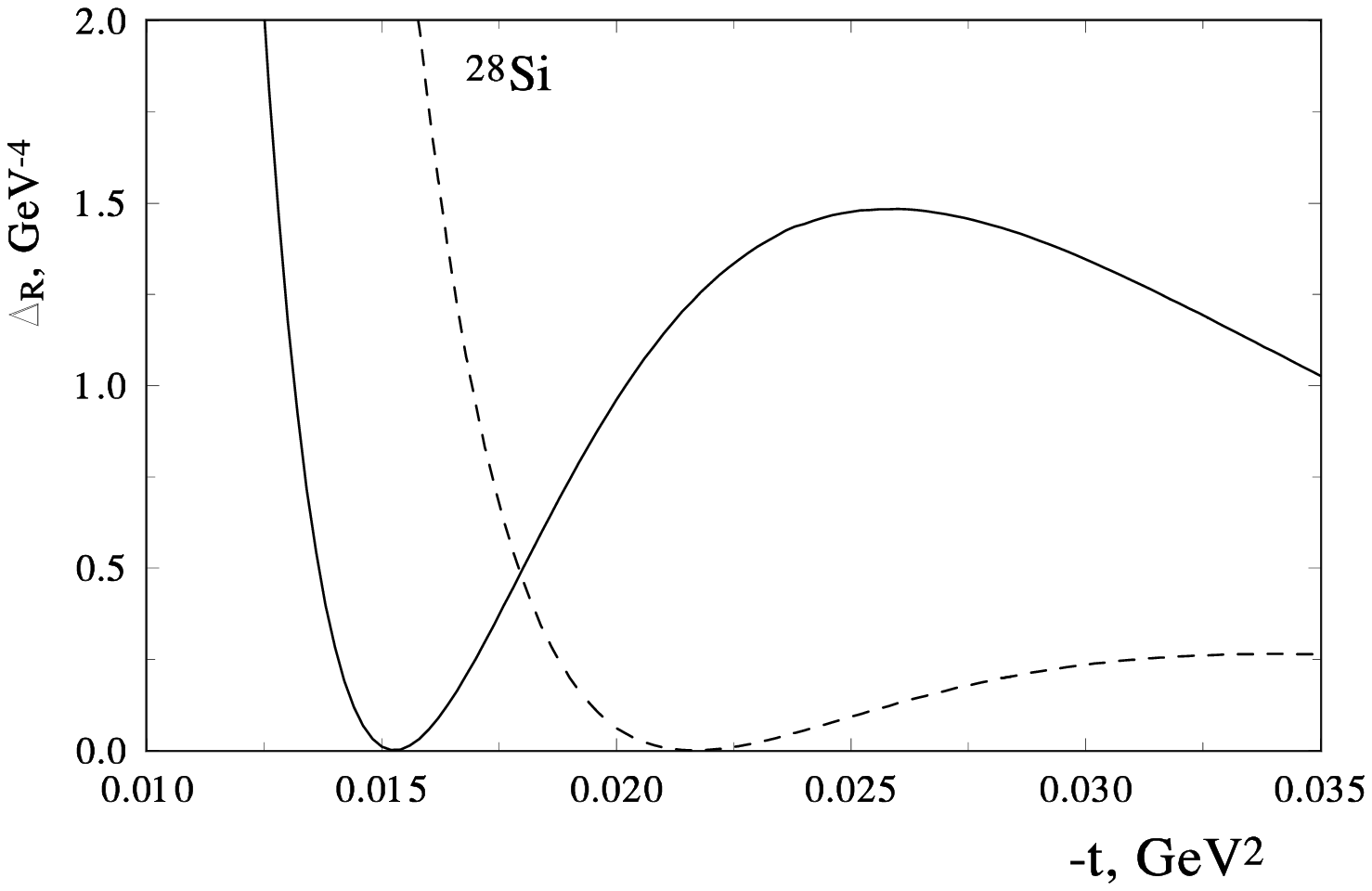}}
\vspace{-0.5cm}
\centerline{Fig.5}
\end{figure}

\newpage

\begin{figure}[!ht]
\epsfysize=80mm
\centerline{\epsfbox{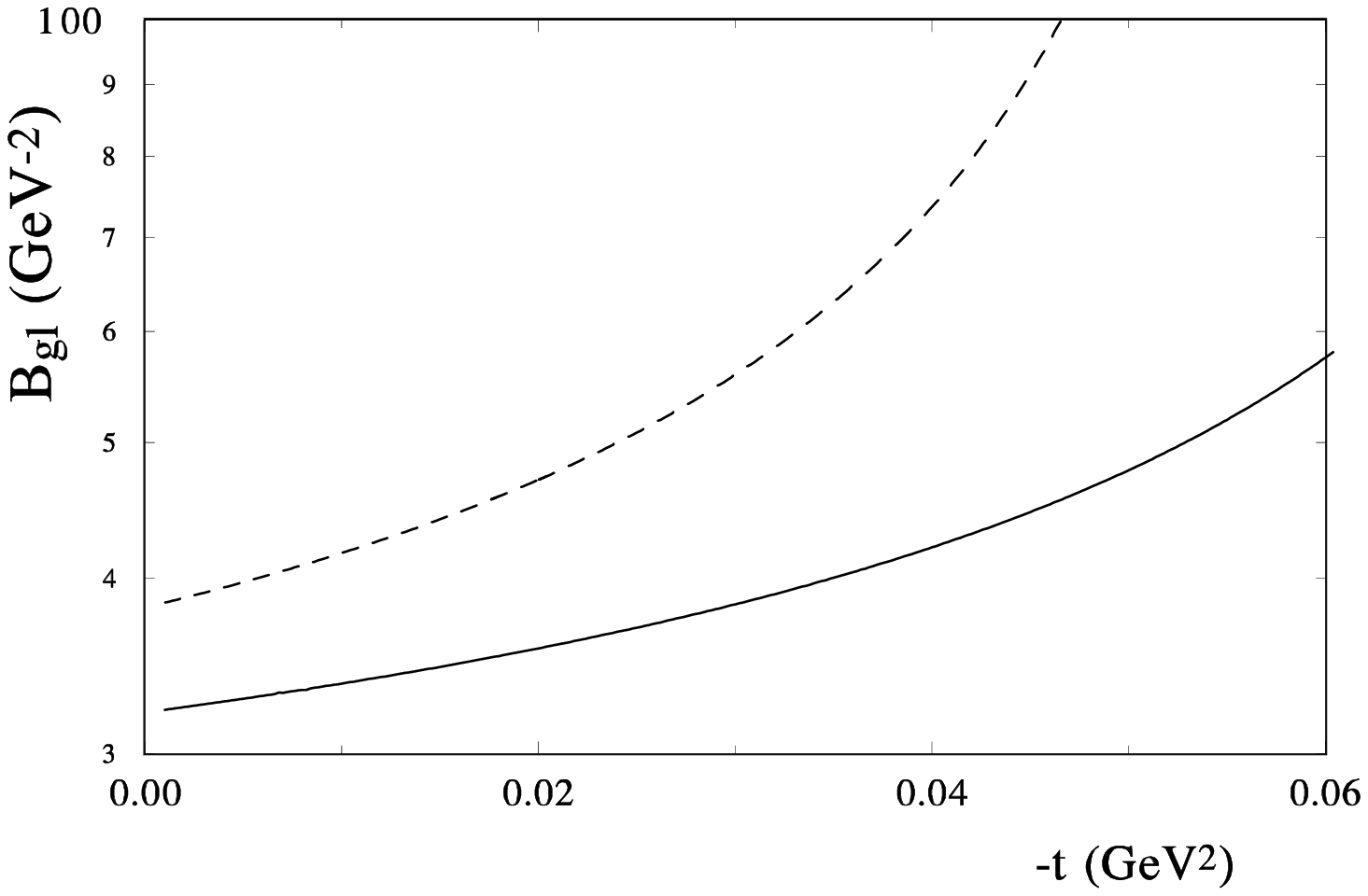}}
\vspace{-0.5cm}
\centerline{Fig.6 }
\epsfysize=80mm
\centerline{\epsfbox{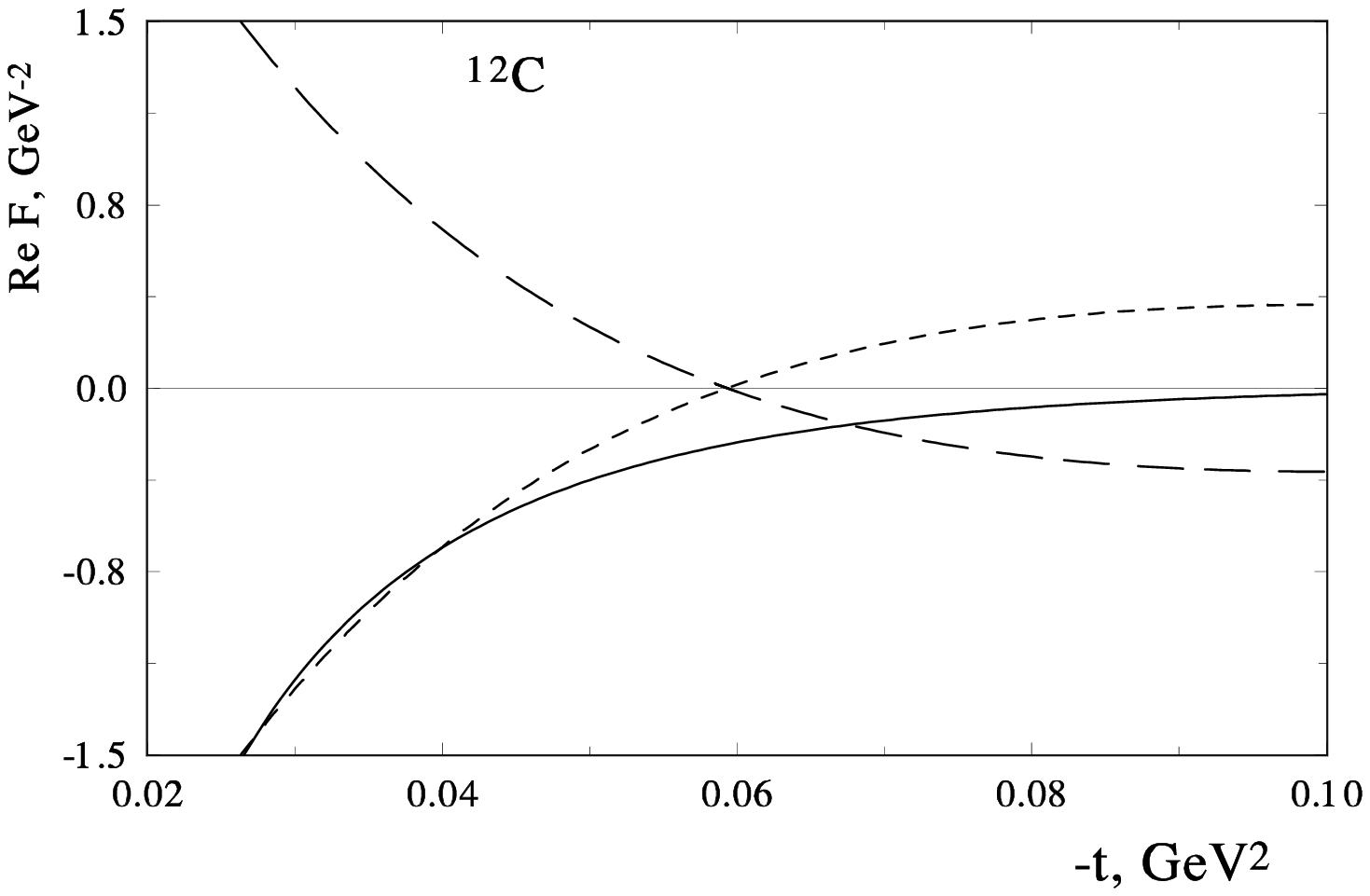}}
\vspace{-0.5cm}
\centerline{Fig.7}
\end{figure}

\newpage

\begin{figure}[!ht]
\epsfysize=70mm
\centerline{\epsfbox{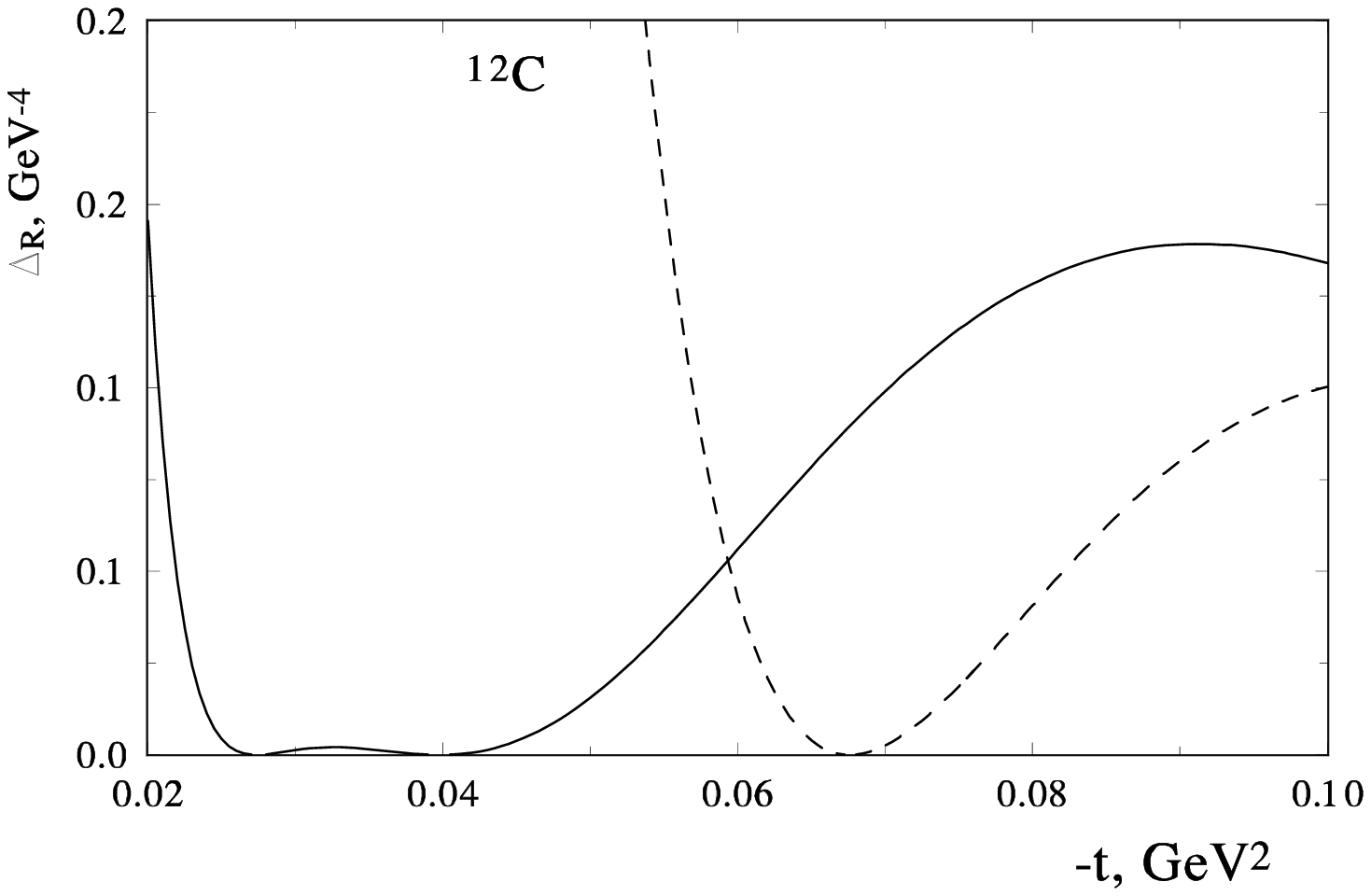}}
\vspace{-0.5cm}
\centerline{Fig.8 a}
\epsfysize=70mm
\centerline{\epsfbox{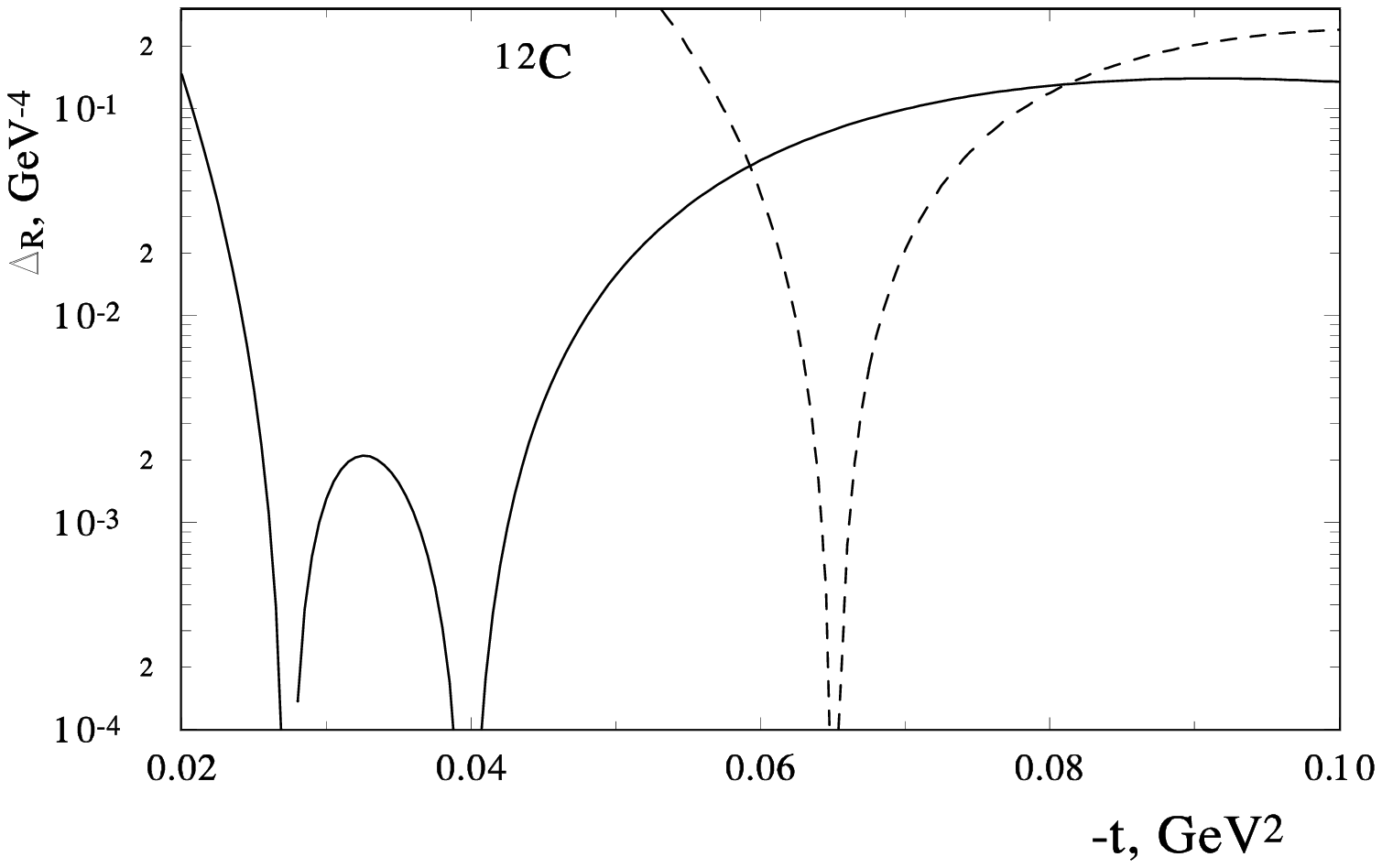}}
\vspace{-0.5cm}
\centerline{Fig.8 b}
\end{figure}

\end{document}